\documentclass[twocolumn,showpacs,fleqn,nobibnotes]{revtex4}

\usepackage{amsmath}
\usepackage{graphicx}
\usepackage{float}
\usepackage{subfigure}

\def\lsim{\raise0.3ex\hbox{$<$\kern-0.75em\raise-1.1ex\hbox{$\sim$}}}
\def\gsim{\raise0.3ex\hbox{$>$\kern-0.75em\raise-1.1ex\hbox{$\sim$}}}

\begin{document}
\newcommand\ie {{\it i.e.}}
\newcommand\eg {{\it e.g.}}
\newcommand\etc{{\it etc.}}
\newcommand\cf {{\it cf.}}
\newcommand\etal {{\it et al.}}
\newcommand{\be}{\begin{eqnarray}}
\newcommand{\ee}{\end{eqnarray}}
\newcommand{\jp}{$ J/ \psi $}
\newcommand{\pp}{$ \psi^{ \prime} $}
\newcommand{\ppp}{$ \psi^{ \prime \prime } $}
\newcommand{\dd}[2]{$ #1 \overline #2 $}
\newcommand\noi {\noindent}

\title{Gluino production in ultrarelativistic heavy ion collisions and nuclear shadowing}
\author{C. Brenner Mariotto $^{a}$, D.B. Espindola $^{b}$ and M.C. Rodriguez
$^{a}$}

\affiliation{
$^a$ Instituto de Matem\'atica, Estat\'{\i}stica e F\'{\i}sica, Universidade Federal do Rio Grande \\
Caixa Postal 474, CEP 96201-900, Rio Grande, RS, Brazil
\\
$^b$ Instituto de F\'{\i}sica, Universidade Federal do Rio Grande do Sul
\\
Caixa Postal 15051, CEP 91501-970,\nolinebreak Porto Alegre, RS, Brazil\\
}

\begin{abstract}
In this article we investigate the influence of nuclear effects in the production of gluinos in nuclear collisions at the LHC, and estimate the transverse momentum dependence of the nuclear ratios 
$R_{pA}  = { \frac{d\sigma (pA)}{dy d^2 p_T} } / A {\frac{d\sigma
(pp)}{dy  d^2 p_T}}$ and $R_{AA}  = { \frac{d\sigma (AA)}{dy d^2 p_T} } / A^2 {\frac{d\sigma
(pp)}{dy d^2 p_T}}$.  We demonstrate that depending on the magnitude of the nuclear effects, the production of gluinos could be enhanced, compared to proton-proton collisions. The study of these observables can be useful to determine the magnitude of the shadowing and antishadowing effects in the nuclear gluon distribution. Moreover, we test different SPS scenarios, corresponding to different soft SUSY breaking mechanisms, and find that the nuclear ratios are strongly dependent on that choice.

\end{abstract}

\maketitle

The main aim of the Large Hadron Collider (LHC), which is already running and soon will be in complete operation with 14 TeV, is to find the Higgs particle. That discovery may either confirm the Standard Model (SM) or open new windows towards new physics. Although the SM explain all experimental data except neutrino masses, there are many reasons to go beyond it. Some theoretical problems in the SM are: hierarchy problem, electroweak symmetry breaking, gauge coupling unification, etc \cite{marcosreview}. The Minimal Supersymmetric Standard Model (MSSM) is the simplest supersymmetric extension of the SM, being a good candidate to Physics Beyond Standard Model \cite{marcosreview,dress}.
In the MSSM, for each usual particle, one assigns a superpartner with oposite statistics: it means that 
for each boson there is a fermionic superpartner, and the reverse in the case of fermions. In the strong sector, one has the so called supersymmetric QCD (sQCD), where besides the gluon (boson) and quarks (fermions), there are the corresponding superpartners: gluinos (fermions) and squarks (bosons). On this model, the gluinos are the superpartners of gluons, they are color octet fermions and therefore they can not mix with other particles, as a result its mass is a parameter of soft SUSY breaking terms. 
Gluinos are Majorana fermions, expected to be one of the most massive MSSM sparticles, and therefore, their 
production is only feasible at very energetic machines such as the LHC. The gluino and squark masses are still unknown parameters, but they cannot be smaller than around a half TeV, as predicted by several models for SUSY breaking. The ``Snowmass Points and Slopes'' (SPS) \cite{sps1} are a set of benchmark points
and parameter lines in the MSSM parameter space corresponding to different scenarios in the search
for supersymmetry at present and future experiments (See \cite{{sps2}} for a very nice review). 
The aim of this convention
is reconstructing the fundamental supersymmetric theory, and its breaking
mechanism, from the experimental data. The different scenarious correspond to three different kinds of models. 
The points SPS 1-6 are Minimal Supergravity (mSUGRA) model, 
SPS 7-8 are gauge-mediated symmetry breaking (GMSB) model, and SPS 9 are 
anomaly-mediated symmetry breaking (mAMSB) model (\cite{sps1,sps2,sps}). Each set of parameters leads to different masses of the gluinos and squarks, which are the only relevant SUSY parameters in our study, and we shown their values in Tab.(\ref{tab:tmasses}). It will be shown below that the choice of SPS scenario affects the results for gluino production.

\begin{table}[t]
\renewcommand{\arraystretch}{1.10}
\begin{center}
\normalsize
 \vspace{0.5cm}
\begin{tabular}{|c|c|c|}
\hline
\hline
Scenario & $m_{\tilde{g}}\, (GeV)$ & $m_{\tilde{q}}\, (GeV)$ \\
\hline
\hline
 SPS1a & 595.2  & 539.9 \\
 SPS1b & 916.1  & 836.2 \\
 SPS2 & 784.4  & 1533.6 \\
 SPS3 & 914.3  & 818.3 \\
 SPS4 & 721.0  & 732.2 \\
 SPS5 & 710.3  & 643.9 \\
 SPS6 & 708.5  & 641.3 \\
 SPS7 & 926.0  & 861.3 \\
 SPS8 & 820.5  & 1081.6 \\
 SPS9 & 1275.2  & 1219.2 \\
  \hline
\hline
\end{tabular}
\caption{The values of the masses of gluinos and squarks in the SPS scenarios.}
\label{tab:tmasses} 
\end{center}
\end{table}

Another aim of the LHC is to study the possible creation and characterization of the so called quark gluon plasma (QGP), which is one of the predictions of the Quantum Chromodynamics (QCD) (see e.g. \cite{reviewqpg}). The heavy ion program at RHIC have brought many interesting results about the evidences of the QGP formation, which is in fact consistent with an almost perfect liquid \cite{rhic}.  
Apart from the QGP, cold matter effects play also a very important role, changing the amount of interacting quarks and gluons in a given kinematic region. 

\begin{widetext}

\begin{figure}[t]
\begin{center}
\includegraphics[width=17.2cm]{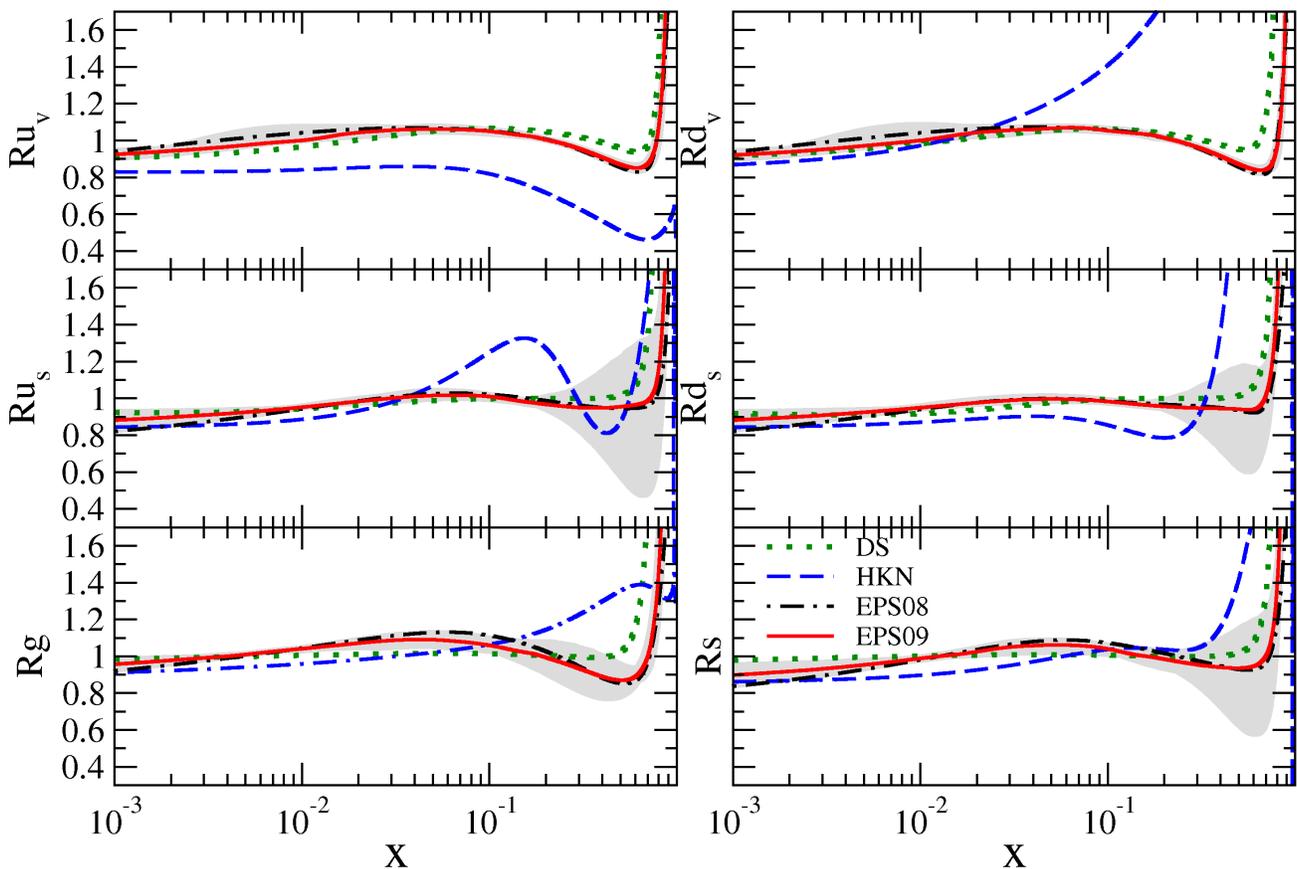}
\caption{\label{rg} (Color online) Ratios $R_f\equiv xf_A/Axf_N$ for the valence, sea quarks and gluons, predicted by the DS \cite{sassot}, HKN \cite{HKN}, EPS08 \cite{EPS08}, and EPS09 \cite{EPS09} parametrizations at $Q=595$ GeV and $A=208$. The uncertainty band is shown for EPS09 nPDF's.}
\end{center}
\end{figure}

\end{widetext}

If the gluinos are found in proton-proton (pp) collisions ($\sqrt{s}=14\, TeV$) at the LHC and if their masses are not much larger than 1 TeV, they might also be produced in collisions involving nuclei - pA (proton-nucleus, $\sqrt{s}=8.8\, TeV$) and AA (nucleus-nucleus, $\sqrt{s}=5.5\, TeV$) LHC modes. In this case, nuclear effects have to be considered in the searches for this supersimmetric particles. One important initial state effect is the so called shadowing effect, which makes the parton distribution functions of the bound proton in a nucleus $A$ (nPDFs) to be different from the usual PDFs in the free proton, $f_i^A(x,Q_0^2)=R_i^A(x,Q_0^2)f_i^p(x,Q_0^2)$, where $R^A_i$ are the nuclear modification ratios which parametrize the nuclear effects. 
There are several parametrizations of nuclear PDFs, based on different assumptions and techniques to perform a global fit of different sets of nuclear experimental data using the DGLAP evolution equations: EKS98 \cite{EKS98}, DS \cite{sassot}, HKN \cite{HKN}, EPS08 \cite{EPS08} and EPS09 \cite{EPS09}, where the two later include different RHIC data for the first time. Also, EPS09 includes an uncertainty band around the central values. The typical $x$ behavior of the nuclear modification ratios is the following: a supression for $x\lesssim 10^{-2}$ (shadowing), followed by an increasement around $10^{-1}$ (antishadowing), again a supression for $x\gtrsim 0.3$ (EMC effect), and a bigger increasement when $x$ approaches 1 (Fermi motion). The whole effect is usually called shadowing.

To ilustrate how shadowing can influence the amount of partons in the nuclear medium, we show in Fig. \ref{rg} 
the results for a few nuclear modification ratios for the gluons ($R_g$), valence ($u_v$, $d_v$) and sea ($u_s$, $d_s$, $s$) quarks. Results for charm and bottom are not shown, since they are not included in some of the parametrizations above (see sec. \ref{gluprod} for more details). The hard scale $Q=595\,GeV$ is the gluino mass (SPS1a scenario, shown in Tab. \ref{tab:tmasses}), which is quite high. We did not include the EKS in our analysis, since this parametrization is not defined for such high $Q$ values. 
Concerning $R_g$, the usual shadowing (suppression) for very low $x$ is present in all parametrizations, being very small for DS ($5\%$ suppression at $x\simeq {10}^{-5}$,  flat behavior), stronger for EPS08 ($25\%$ suppression at $x\simeq {10}^{-5}$) and moderate for HKN and EPS09 ($15\%$ suppression at $x\simeq {10}^{-5}$). However, for the processes considered in this work, the small-$x$ region do not contribute (see below), and therefore we only show the relevant $x$ domain. The shadowing is very smaller for $x\gtrsim 10^{-3}$, the DS and EPS being inside the EPS error band in many $x$ regions (except for very high $x$).  
On the other hand, at larger $x$, antishadowing (enhancement) is present in EPS08, EPS09 ($x\le {10}^{-1}$) and HKN (larger $x$), but not in DS. The behavior with increasing $x$ is also different, being the growth steeper for EPS08, and smoothed out in later EPS09. 
Concerning the other parton species, $R_{u_V} \sim R_{d_V}$ and $R_{u_S} \sim R_{d_S}$ for all parameterizations except for HKN which show rather large differences.
For moderate values of $x$, the HKN valence $d$ and gluons are enhanced, the valence $u$ is suppressed, while the sea HKN $u$ has an enhancement followed by a suppression at larger $x$ (EMC effect). 
There are many investigations on inclusive heavy quark, quarkonium and prompt photon production in central proton-nucleus and nucleus-nucleus collisions (See e.g. Refs. 
\cite{vogt1,vogt2,vogt3,vogt4,vic_luiz1,ABMG:2006,BMG:2008,DMM:2009, BMM:2009}), which try to help in constraining the nuclear PDFs from several observables. The variety of nuclear effects may also be relevant for gluino production, since there are contributing diagrams with both (anti)quarks and gluons in the initial state.

In the case of gluino production, because of the large gluino masses, the values of $x$ probed tend to be quite high (from $x\gtrsim 10^{-2}$ to almost $1$), 
and then the antishadowing, EMC effect and even Fermi motion may be important (depending on the kinematic region and nuclear PDF), which may enhance the gluino production rate compared to that obtained from single nucleon collisions at the same energy. Therefore, whereas the smaller center of mass energy (5.5 TeV (AA) and 8.8 TeV (pA)) will reduce the gluino production rates (compared to 14 TeV (pp)), there may be an enhancement due to the amount of quarks and gluons on the nuclear medium compared with the nucleon parton distributions on a single proton, due to high density nuclear effects. In this work we investigate whether this enhancement/suppression is present or not.

This article is organized as follows. The basic formulae to calculate gluino production are presented in section \ref{gluprod}. Our results for gluino produced in nuclear collisions at the LHC are presented in section \ref{nuclglu}, followed the the conclusions.

\section{Gluino production in pp colisions \label{gluprod}}

In order to make a consistent comparison and for sake of simplicity, we restrict ourselves to leading-order
(LO) accuracy, where the partonic cross-sections for the production of squarks and gluinos in hadron
collisions were calculated at the Born level already quite some time ago \cite{Dawson}. 
The corresponding NLO calculation has already been done for the MSSM case \cite{Zerwas}, and the
impact of the higher order terms is mainly on the normalization of the cross
section \cite{Zerwas}, which cancels out in the ratios.

\begin{figure}[tb]
\begin{center}
\includegraphics[scale=0.79]{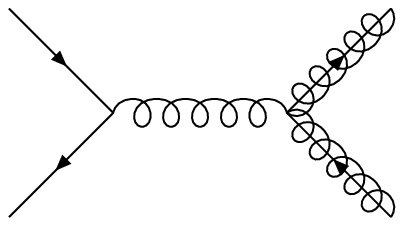}\hspace{0.2cm}
\includegraphics[scale=0.75]{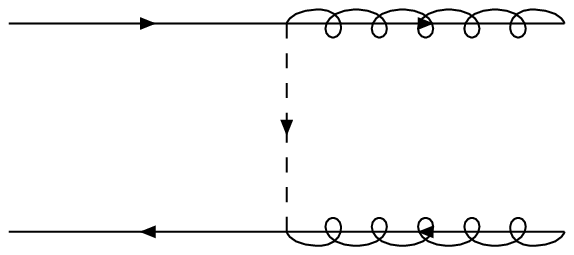}
\\ \vspace{0.2cm}(a)\\\vspace{0.2cm}
\includegraphics[scale=0.79]{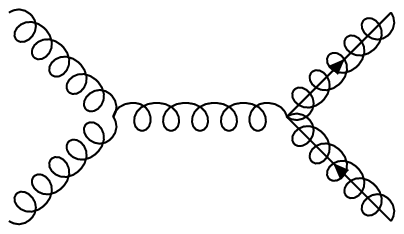}\hspace{0.2cm}
\includegraphics[scale=0.75]{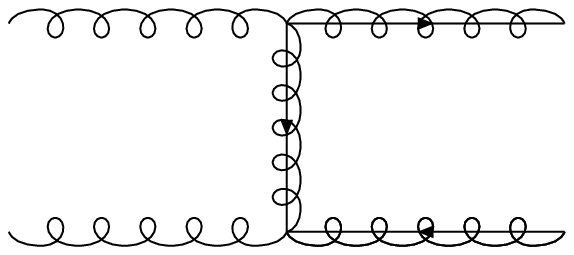}
\\ \vspace{0.2cm}(b)\\\vspace{0.2cm}
\includegraphics[scale=0.79]{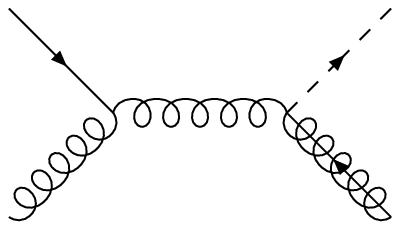}\hspace{0.2cm}
\includegraphics[scale=0.75]{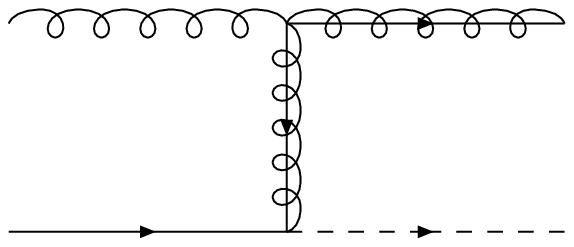}
\\ \vspace{0.2cm}(c)
\end{center}
\caption{Lowest order Feynman diagrams for gluino production: (a) quark-antiquark initial states and (b) gluon-gluon initial states (double gluino production); (c) quark-gluon initial states (squark--gluino production).}
\label{fig:GG}
\end{figure}

The contributing LO diagrams for inclusive gluino production in proton-proton collisions are $g g  \rightarrow \tilde{g} \tilde{g}$, $q \bar q  \rightarrow \tilde{g} \tilde{q}$ and the Compton process $g q  \rightarrow \tilde{g}
\tilde{q}$, (shown in Fig. ~\ref{fig:GG}), where one has to be carefull in including Feynman rules for Majorana particles \cite{sbf}. 

Incoming quarks (including incoming $b$ quarks) are assumed to be massless,
such that we have $n_f=5$ light flavours. We only consider final state 
squarks corresponding to the light quark flavours. All 
squark masses are taken equal to $m_{\tilde q}$ 
($L$-squarks and $R$-squarks are therefore mass-degenerate  
and experimentally indistinguishable.) 
We do not consider in detail top squark production where these assumptions do not hold and which require
a more dedicated treatment~\cite{plehn}.

The invariant cross section for single gluino production can be written as \cite{Dawson}
 \begin{eqnarray}
E\frac{d\sigma }{d^3p}= \sum_{ijd} \int_{x_{min}}^1 dx_a
f_i^{(a)}(x_a,\mu )f_j^{(b)}(x_b,\mu )\nonumber \\
\frac{x_ax_b}{x_a-x_{\perp} \left( \frac{\zeta + \cos \theta}{2\sin \theta} \right) }
\frac{d \hat{\sigma}}{d \hat{t}}(ij\rightarrow \tilde{g} d), 
\label{singlegluino}
\end{eqnarray}
where $f_{i,j}$ are the parton distributions of the incoming protons 
and $\frac{d \hat{\sigma}}{d \hat{t}}$ is the LO partonic cross section
\cite{Dawson} for the subprocesses involved. The identified gluino is produced at center-of-mass angle $\theta$ and 
transverse momentum $p_T$, and $x_{\perp}=\frac{2p_T}{\sqrt{s}}$. The Mandelstam variables  
of the partonic reactions $ij\rightarrow \tilde{g}\tilde{g}, \tilde{g}\tilde{q}$ 
are then
\begin{eqnarray}
\hat{s}&=& x_ax_bs, \nonumber \\ 
\hat{t}&=& m_{\tilde{g}}^2-x_ax_{\perp}s\left( \frac{\zeta-\cos \theta}{2\sin \theta}\right), 
\nonumber \\ 
\hat{u}&=& m_{\tilde{g}}^2-x_bx_{\perp}s\left(\frac{\zeta+\cos \theta}{2\sin \theta}\right).
\end{eqnarray}
Here
\begin{eqnarray}
x_b&=&\frac{2\upsilon + x_ax_{\perp}s\left( \frac{\zeta-\cos \theta}{\sin \theta}\right)
}{2x_as-x_{\perp}s\left(\frac{\zeta+\cos \theta}{\sin \theta}\right) },
\nonumber \\ 
x_{min}&=&\frac{2\upsilon + x_{\perp}s\left(\frac{\zeta+\cos \theta}{\sin \theta}\right) }
{2s-x_{\perp}s\left( \frac{\zeta-\cos \theta}{\sin \theta}\right) }, \nonumber \\
\zeta &=& {\left( 1+\frac{4m_{\tilde{g}}^2\sin ^2
\theta}{x_\perp^2s} \right)}^{1/2}, \nonumber \\ 
\upsilon &=& m_d^2-m_{\tilde{g}}^{2}, 
\end{eqnarray}
where $m_{\tilde{g}}$ and $m_d$ are the masses of the final-state partons
produced. 
The center-of-mass angle $\theta$ and the differential cross section above
can be easilly written in terms of the pseudorapidity variable $\eta=-\ln \tan
(\theta/2)$,
 which is one of the experimental observables.

Predictions for gluino production in $pp$ collisions at the LHC ($\sqrt{s}=14$ TeV), in all SPS scenarious, are shown in a former work \cite{MariottoRodriguez}, where there is a huge difference in the magnitude of $p_T$ distributions for different SPS points, making it possible to distinguish between some different SUSY breaking scenarios. We can ask if the same occurs in nuclear processes, and answering this question is also a goal on this article.

\section{Gluino production in nuclear collisions \label{nuclglu}}

Let us now focus on gluino production in nuclear collisions. The calculation is done as explained in the previous section, replacing the parton distributions in the free nucleon ($f_i^p$ in Eq. (\ref{singlegluino})) by the corresponding nuclear parton distributions $f_i^A$ (for the proton PDF we use the CTEQ6L1 \cite{Pumplin:2002vw}). The nuclear effects are then studied by comparing the different nPDF's available (for consistency, we use the LO version of all nPDF's). To be sure that the nPDF's are within the regions of validity, we have used $Q=m_{\tilde{g}}$ as the hard scale (as done in \cite{dress}). Another possible choice, a $p_T$ running $Q=m_{\tilde{g}}+p_T$ scale, would push some of the nPDF's outside the region of valididy (EPS08 and EPS09 are frozen in $Q=1000 \,GeV$ for values above that scale, whereas DS is not valid in that region). For this reason, the DS could not be considered in the SPS9 scenario (see Table I), with extra large gluino masses. To start with, we consider the SPS1a scenario as the first (most optimistic) choice of gluino and squark masses. 

\begin{figure}[t]
 \includegraphics[width=8.6cm]{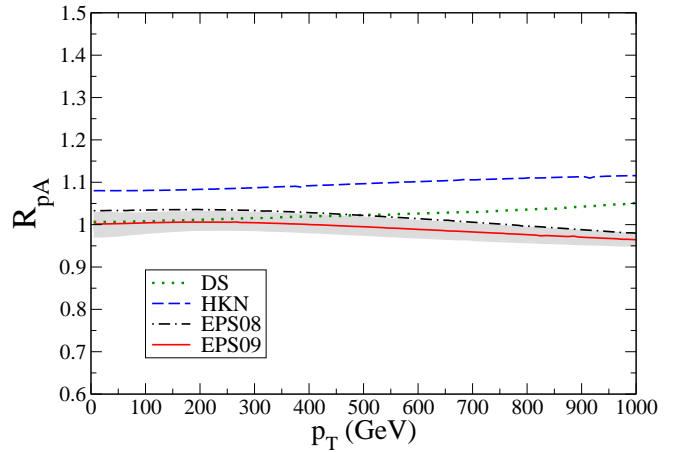}
\caption{\label{gluinopA} (Color online) Transverse momentum dependence of the nuclear modification factor $R_{pA}$ for inclusive gluino production in pA collisions at the LHC ($\sqrt{s}=8.8$ TeV, $|\eta|\le 2.5$), for distict nPDFs.}
\end{figure}

In Fig. \ref{gluinopA} we show our results for the transverse momentum dependence of the nuclear modification factor defined by 
\begin{eqnarray}
R_{pA}  \equiv  \frac{d^2\sigma (pA)}{d\eta d p_T} / A \frac{d^2\sigma
(pp)}{d\eta d p_T}\,\,,
\end{eqnarray}
for gluino production in proton-nucleus collisions at the LHC ($\sqrt{s}=8.8$ TeV). For lower $p_T$, the DS and EPS08 nPDF's are inside the EPS09 uncertainty band, with almost no nuclear effect, $R_{pA}\sim 1$. For $p_T>500\,GeV$, the EPS's starts to be slightely suppressed (increasing with $p_T$), whereas the DS starts to be slightely enhanced (increasing with $p_T$). For the HKN distribution, there is a larger enhancement of 10$\%$, increasing slowly with $p_T$. This means that the correct amount of (anti)shadowing is undefined. In fact, as $p_T$ grows the probed values of $x$ increase, and the EPS nPDF's enter the EMC region, whereas this effect does not appear for the other nPDF's.

\begin{figure}[t]
  \includegraphics[width=8.6cm]{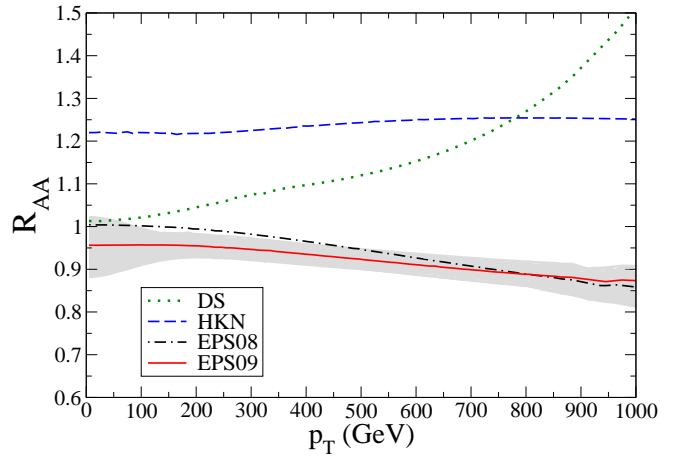}
\caption{\label{gluinoAA} (Color online) Transverse momentum dependence of the nuclear modification factor $R_{AA}$ for inclusive gluino production in AA collisions at the LHC ($\sqrt{s}=5.5$ TeV, $|\eta|\le 2.5$), for distict nPDFs.}
\end{figure}

In Fig. \ref{gluinoAA} we present a similar analysis for the transverse momentum dependence of the nuclear modification factor defined by
\begin{eqnarray}
R_{AA}  =  \frac{d^2\sigma (AA)}{d\eta d p_T} / A^2 \frac{d^2\sigma
(pp)}{d\eta d p_T}\,\,,
\end{eqnarray}
for gluino production in nucleus-nucleus collisions at the LHC ($\sqrt{s}=5.5$ TeV). In this case, the nuclear effects are amplified because of the presence of two nuclei. Besides, the probed values of $x$ are pushed into very high-$x$ due to the smaller center of mass energy. Indeed, the EPS supression increases with $p_T$ in a stronger way than in the pA case (around $15\%$ for higher $p_T$). The DS nPDF has an enhancement pattern, increasing with $p_T$, which shows that this distribution has reached the Fermi motion effect in the very right side of Fig. 1. The enhancement is also larger for the HKN, above $20\%$ with a very tiny increasement with $p_T$. It seems that, if the latest EPS09 nPDF is the more correct distribution, the gluino production will be slightely suppressed compared with $pp$ collisions at the same energy, whereas the DS and HKN suggests that there will be some enhanced production of gluinos in nuclear collisions.

The possible increasement of the gluino production rate in nuclear collisions (compared with pp collisions at same energy) shown above is in fact too low to really improve the small feasibility of detecting the gluinos when going from pp to pA and AA. In fact, the higher hadronic activity in nuclear collisions make the detection of gluinos more difficult, and the smaller CM energy available produces a smaller number of gluinos compared to 14 TeV pp colisions.  The expected luminosity to be reached in the AA collisions (${\cal{L}}_{NN}\approx 10^{27}\,A^2\,cm^{-1}s^{-1}$) \cite{Jowett:2008hb} is seven orders of magnitude smaller than in the pp mode (${\cal{L}}_{pp}\approx 10^{34}\,cm^{-1}s^{-1}$), and this is the main limitation to detecting nuclear gluinos (they will be produced but will hardly be seen). In the pA mode, one expects a luminosity of ${\cal{L}}_{pA}\approx 7.4\times 10^{29}\,cm^{-1}s^{-1}$ \cite{Klein:2002wm}, which becomes $7.4\,pb^{-1}$ assuming a full LHC year $10^7\,s$ (one usually considers a month ion running time $10^6\,s$) in the ion mode. With only our LO estimation, and considering the more suppressed EPS09, one would than obtain around 31 gluinos produced in the pA mode for the $p_T$ integrated region, so statistics is really limited. It has been suggested that the pA luminosity could eventually be upgraded to ${\cal{L}}_{pA}\approx 10^{31}\,cm^{-2}s^{-1}$ \cite{d'Enterria:2009er}, in this case our estimate would increase to 430 gluinos in one year run. For more realistic estimates, the NLO correction would still increase the cross-sections for the various production processes by up to a factor of less than two \cite{Zerwas}.

\begin{widetext}

\begin{figure}[t]
\begin{center}
\includegraphics[width=17.2cm]{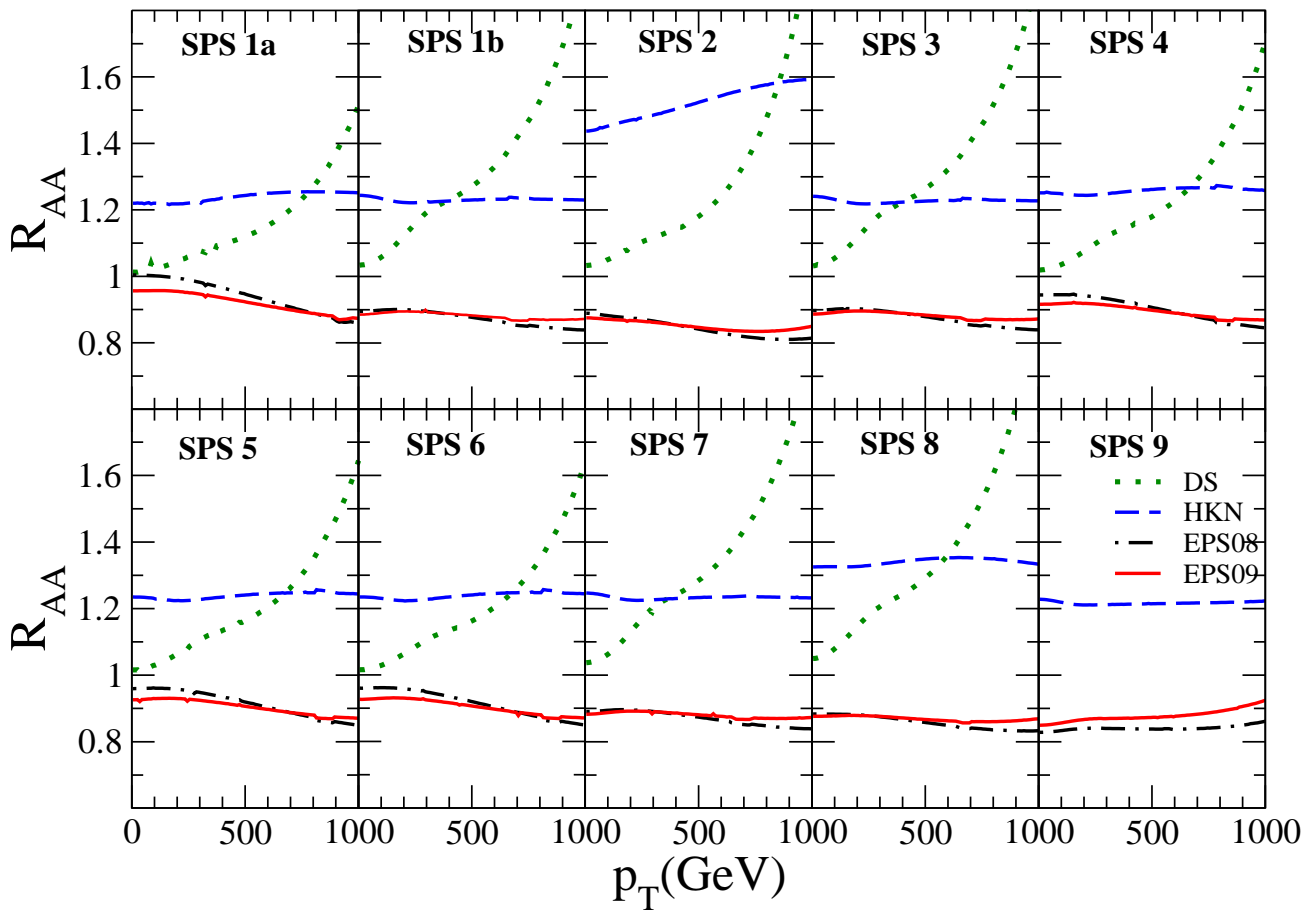}
\caption{\label{gluinoAAsps} (Color online) Transverse momentum dependence of the ratio $R_{AA}$ in single gluino production at the LHC ($\sqrt{s}=5.5\, TeV$), for different choices of nuclear parton distributions: DS \cite{sassot}, HKN \cite{HKN} EPS08 \cite{EPS08} and EPS09 \cite{EPS09}, in different SPS scenarios.}
\end{center}
\end{figure}

\end{widetext}

Not only the nuclear shadowing but also the SUSY breaking parameters affect the nuclear ratios. This dependence is indirect, since the gluino and squark masses ($m_{\tilde{g}}$, $m_{\tilde{q}}$), are the only parameters that really affect the results, but these masses are consequences of the different SUSY breaking parameters in the different SPS scenarious. 
This is shown in Fig. \ref{gluinoAAsps}, where different SPS scenarious give different absolute values for the $R_{AA}$ nuclear ratios (this can be seen by comparing for example the starting point of each curve). The $p_T$ growth for the DS nPDF is even more steeper for the higher mass SPS scenarious (higher $x$). For the SPS9 scenario, the results should not be trusted, since most parametrizations are not valid in that region: 
the HKN predicts an enhancement essentially constant with $p_T$, and the frozen EPS's supression decreases with $p_T$. Because of the odd interplay of nuclear effects and SUSY breaking scenarios, one needs to put better constraints on the nuclear PDF's before describing precisely gluino production in nuclear collisions. Conversely, the discovery and measurement of the gluino and squark masses will be important in the searches for sparticles produced in nuclear collisions, taking into account the correct nuclear effects which also depend on the those masses.

\section{Conclusions}
To conclude, in this work we studied the nuclear effects in pA and AA gluino production at the LHC. 
We have shown different results of enhancement or suppression depending on the nuclear PDF, the effects being smaller in pA interactions and larger in nuclei collisions. Gluinos will probably be copiously produced in the $pp$ channel. Once the details
of gluino production are known in $pp$ interactions, studying this
final state in $pA$ and $AA$ collisions could give unprecedented
constraints on the nPDFs in a heretofore unexplored region of $Q^2$. One could use the higher energy to get a good
measurement of gluino production and search for deviations from that
in the measurable $p_T$ range for $pA$ and $AA$ to measure quark and
gluon shadowing at very high scales where nothing at all is known
about it. Uncertainties on the nPDF's (and cold matter effects in general), and on the SUSY breaking scenarios (which give different masses for the gluinos and squarks) has to be disentangled in the future searches. 
For heavy nuclei collisions, where its expected the formation of the quark gluon plasma, it may appear other channels where gluino is produced. Here we only investigated cold matter effects, namely the shadowing of the nuclear distributions. If gluinos are discovered in pp collisions at LHC, they will also be there for pA and AA. However, the ability to search for them will depend on a further understanding of the correct nuclear effects. 

\begin{acknowledgments}
This work was partially financed by the Brazilian funding agency
CNPq. 
\end{acknowledgments}

\end{document}